\documentclass{article}
\usepackage{amssymb, amsmath}
\numberwithin{equation}{section}
\newtheorem{theorem}{Theorem}[section]
\newtheorem{proposition}{Proposition}[section]

\newtheorem{remark}{Remark}[section]
\newtheorem{definition}{Definition}[section]
\begin{document}
\title{Global Existence of Solutions for the Relativistic Boltzmann Equation with Arbitrarily Large Initial Data  on a Bianchi Type I Space-Time}
\author{Norbert NOUTCHEGUEME$^{1}$; David DONGO$^{1}$ ; Etienne TAKOU$^{1}$\\
$^{1}$Department of Mathematics, Faculty of Sciences, University
of Yaounde 1,\\ PO Box 812, Yaounde, Cameroon \\ {e-mail:
nnoutch@justice.com,  nnoutch@uycdc.uninet.cm }}
\date{}
\maketitle
\begin{abstract}
We prove, for the relativistic Boltzmann equation on a Bianchi
Type I space-time, a global existence and uniqueness theorem, for
arbitrarily large initial data.
\end{abstract}
\section{Introduction}
The Boltzmann equation is one of the basic equations of the
kinetic theory. This equation rules the dynamics of a kind of
particles subject to mutual collisions, by determining their
distribution function, which is a non-negative real-valued
function of both the position and the momentum of the particles.
the distribution function, also called the ''phase space density''
of the particles, is physically interpreted as ''the probability
of the presence density'' of the particles in a \\ given volume,
during their  collisinal evolution, and is an essential tool for
such a statistical description.

 In the case of instantaneous, localized, binary and elastic
 collisions, the distribution function is determined by the
 Boltzmann equation, through a non-linear operator called the
 ''collision operator'', that acts only on the momentum of the
 particles, and that describes, at any time, at each point where
 two particles collide with each other, the effects of the
 behavior imposed by the collision to the distribution function,
 taking into account the fact that the momentum of each particle
 is not the same, before and after the collision, only the sum of their two momenta being
 preserved.

 Due to its importance in the kinetic theory, several authors
 studied and proved local and global in time existence theorems
 for the Boltzmann equation, in both the
 \textbf{non-relativistic} case, that considers particles with
 low velocities, and the \textbf{full-relativistic} case, which
 includes the case of fast moving particles with arbitrarily high
 velocities, such as, for example, particles of ionized gas in
 some medias at very high temperature as : burning reactors, solar
 winds, nebular galaxies.
\begin{itemize}
\item [1)]In the non-relativistic case, the original global result
is due to Carleman, T. , in \cite{carleman} ; Diperna, R. J. and
Lions, P. L. proved global existence and weak stability in
\cite{diperna}. Di Blasio, G. , proved the differentiability of
spatially Homogeneous Solutions of the Boltzmann Equation in the
non Maxwellian case in \cite{blasio}. Illner, R. , and Shinbrot,
M. , proved a global result in \cite{illner}, in the case of small
initial data and without symmetry assumption ; an analogous result
is unknown in the full relativistic case.
\item[2)]In the full relativistic case, several authors proved
local existence theorems, considering this equation alone, as
Bichteler, K. , in \cite{bichteler}, Bancel, D. , in
\cite{bancel1}, or coupling it to other fields equations as
Bancel, D. , and Choquet-Bruhat, Y. , in \cite{bancel2}. Glassey,
R. , T. , and Strauss, W. , obtained a global result in
\cite{glassey1}, in the case of data near to that of an
equilibrium solution with non-zero density. Noutchegueme, N. and
Tetsadjio, E. M, proved Global existence for small initial data on
the Minkowski space-time in \cite{noutch}.
\\The objective of this paper is to extend to the full
relativistic case, the global existence theorem in the case of
arbitraliry large  initial data. That was certainly one of the
goals of Mucha, P. , B. , who studies in \cite{mucha1} and
\cite{mucha2}, the relativistic Boltzmann equation, coupled to
Einstein's equations, for a flat Robertson-Walker space-time.
Unfortunately, several points in that work are far from clear,
such as : using a formulation which is valid only in the
non-relativistic case .
\end{itemize}
In this paper, we begin, first of all, by giving, following
Glassey, R. , T. , in \cite{glassey2}, the correct formulation of
the relativistic Boltzmann equation, taking as back-ground
space-time a Bianchi type I space-time . We consider the
homogeneous case, which means that the distribution function only
depends on the time and the momentum of the particles. Such cases
are very useful for instance in cosmology. Next, we build a
suitable functional frame-work, and we construct a sequence of
operators to approximate the ''collision operator''; this provides
a sequence of real-valued functions that converges to a global
solution of the Boltzmann equation. We prove this way, using very
simple a priori estimates that lead to a considerable
simplification of the method followed in \cite{mucha1}, that, if
the coefficients of the back-ground metric are bounded away from
zero, then the initial value problem for the relativistic
Boltzmann equation on that Bianchi type I space-time, has a unique
global solution, for arbitraliry large  initial data and that the
solution admits a very simple estimation by the initial data.

The paper is organized as follows : In section 2, we specify the
notations, we define the function spaces, we introduce the
relativistic Boltzmann equation and the ''collision operator''. We
end this section by a sketch of the strategy adopted. In section
3, we give the properties of the ''collision operator'', some
preliminary results, and we construct the approximating operators.
Section 4 is devoted to the global existence theorem.
\section{Notations, function spaces and the relativistic Boltzmann equation}
\subsection{Notations and function spaces}
A Greck index varies from $0$ to $3$ and a latin index from $1$ to
$3$ . We consider as back-ground space-time, a Bianchi type I
space-time denoted $(\mathbb{R}^{4}, g )$, where, for $x =
(x^{\alpha}) = (x^{o}, x^{i}) \in \mathbb{R}^{4}, x^{o} = t$,
denotes the time and $\bar{x} = (x^{i})$ the space ; $g$ stands
for the metric tensor with signature $(-, +, +, +)$ that can be
written :
\begin{equation}\label{eq:2.1}
g = -dt^{2} + a^{2}(t)(dx^{1})^{2} + b^{2}(t)[(dx^{2})^{2} +
(dx^{3})^{2}]
\end{equation}
in which, $a(t)$ and $b(t)$ are given non-negative regular,
real-valued functions. We require that $a(t)$ and $b(t)$ are
bounded from below i.e. :
\begin{equation}\label{eq:2.2}
a(t) > C_{o},\quad b(t) > C_{o}
\end{equation}
where $C_{o} > 0$ is a given constant. Let us obseve that for $a =
b $, $(\mathbb{R}^{4}, g)$ reduces to the fat Robertson-Walker
space-time, which is the basic model for the study of the
expanding Universe. We consider the collisional evolution of a
kind of uncharged particles in the time-oriented curved space-time
$(\mathbb{R}^{4}, g).$ An essential tool to describe the dynamics
of such particles is their distribution function we denote by $f$,
and that is a non-negative real-valued function of both the
position $(x^{\alpha})$, the 4-momentum $p = (p^{\alpha}) =
(p^{o}, p^{i})$ of the particles, and that coordinalize the
tangent bundle $T(\mathbb{R}^{4})$ i.e :
\begin{equation*}
  f: T(\mathbb{R}^{4}) \simeq
  \mathbb{R}^{4}\times\mathbb{R}^{4}\rightarrow\mathbb{R}^{+},\quad
(x^{\alpha}, p^{\alpha})\longmapsto f(x^{\alpha}\ , p^{\alpha})
\in\mathbb{R}^{+}.
\end{equation*}
We define a scalar (or dot) product and a norm on
$\mathbb{R}^{3}$, by setting, for \\$p = (p^{o}, p^{i}) = (p^{o},
\bar{p})$ and $q = (q^{o}, q^{i}) = (q^{o}, \bar{q})$ :
\begin{equation}\label{eq:2.3}
\bar{p}.\bar{q} = a^{2}(t)p^{1} q^{1} + b^{2}(t)({p}^{2}{q}^{2} +
p^{3}q^{3});  \quad  \mid \bar{p} \mid^{2} =a^{2}(t)(p^{1})^{2} +
b^{2}(t)[(p^{2})^{2}+(p^{3})^{2}]
\end{equation}
We consider massive particles with a rest mass that can be
rescaled to $m = 1$. The particles are then required to move on
the future sheet of the mass-shell whose equation is $g(p, p) =
-1$, or equivalently, using (\ref{eq:2.1}) and (\ref{eq:2.3}):
\begin{equation}\label{eq:2.4}
p^{o} = \sqrt{1 + \mid \bar{p} \mid^{2}}
\end{equation}
The relation (\ref{eq:2.4}) shows that $f$ is in fact  defined on
the subbundle of $T(\mathbb{R}^{4})$ whose local coordinates are
$x^{\alpha}$ and $p^{i}$. \\
Now we consider the homogeneous case for which $f$ depends only on
the time $x^{o} = t$ and $\bar{p} = (p^{i})$. The framework we
will refer to is $L^{1}(\mathbb{R}^{3})$ whose norm is denoted
$\parallel .\parallel$ ; we set, for  $r\in\mathbb{R}, r>0 $ :
\begin{equation*}
X_{r} = \{ f \in L^{1}(\mathbb{R}^{3}),\quad f \geq 0,\quad a.e. ,
\, \parallel  f \parallel \, \leq \, r \}.
\end{equation*}
Endowed with the distance induced by the norm
$\parallel.\parallel$, $X_{r}$ is a complete and connected metric
space. Let I be a real-interval; we denote by
$C[I;L^{1}(\mathbb{R}^{3})]$ the Banach space of continuous and
bounded functions $g: I \longmapsto L^{1}(\mathbb{R}^{3}) $,
endowed with the norm $\parallel \mid g \parallel \mid =
\underset{t \in I}{Sup} \parallel g(t) \parallel$. We set :
\begin{equation*}
  C[I; X_{r}] = \{ \, g\in C[I;L^{1}(\mathbb{R}^{3})], \, g(t)\in X_{r}\, , \, \forall t \in I \}
\end{equation*}
Endowed with the distance induced by the norm $\parallel\mid .
\parallel\mid$, $C[I; X_{r}]$ is a complete metric space.
\subsection{The relativistic Boltzmann equation}
The relativistic Boltzmann equation for uncharged particles, in
the homogeneous case, on the Bianchi type I space-time
$(\mathbb{R}^{4}, g)$ we consider , can be written :
\begin{equation}\label{eq:2.5}
\frac{\partial f}{\partial t} -2 \,\frac{\dot{a}}{a} \,
p^{1}\frac{\partial f}{\partial p^{1}}
 -2 \, \frac{\dot{b}}{b} \, p^{2}\frac{\partial f}{\partial p^{2}} - 2 \, \frac{\dot{b}}{b} \, p^{3}\frac{\partial f}{\partial p^{3}}
 =\frac{1}{p^{o}}\, Q (f, f)
\end{equation}
with $\dot{a} = \frac{da}{dt}$, and where $Q$ is the collision
operator we now introduce. In the case of instantaneous,
localized, binary and elastic collisions we consider, the
collision operator $Q$ that acts only on the momentum variable, is
defined as follows, $p$ and $q$ standing for the momenta of $2$
particles before their collision, $p'$ and $q'$ standing for their
momenta after the collision, regardless for the time $t$, and
where $f$ and $g$ are $2$ functions on $\mathbb{R}^{3}$ :
\begin{equation}\label{eq:2.6}
1) \qquad\qquad\qquad \qquad\qquad\qquad\qquad\qquad   Q(f, g) =
Q^{+}(f, g) - Q^{-}(f, g)
\end{equation}  where
\begin{equation}\label{eq:2.7}
  2)\qquad\qquad  Q^{+}(f, g) =
\int_{\mathbb{R}^{3}}\frac{a(t)b^{2}(t)}{q^{o}}d\bar{q}\int_{S^{2}}f(\bar{p'})g(\bar{q'})S(a(t),
b(t) , \bar{p}, \bar{q},\bar{p'},\bar{q'} )d\omega
\end{equation}
\begin{equation}\label{2.8}
 3)\qquad \,\,\qquad Q^{-}(f, g) =
\int_{\mathbb{R}^{3}}\frac{a(t)b^{2}(t)}{q^{o}}d\bar{q}\int_{S^{2}}f(\bar{p})g(\bar{q})S(a(t),
b(t), \bar{p}, \bar{q},\bar{p'} ,\bar{q'} )d\omega
\end{equation}
in which
 \\ 4) $S^{2}$ is the unit sphere of $\mathbb{R}^{3}$
whose area element is denoted $d\omega$ ;\\
 5) $S$ is a
non-negative real-valued function of the indicated arguments,
called the collision kernel or the cross section of the
collisions, and for which we require the boundedness and the
symmetry assumptions :
\begin{equation}\label{eq:2.9}
0 \leq S(a(t), b(t),\bar{p},\bar{q},\bar{p'},\bar{q'}) \leq C_{1}
\end{equation}
\begin{equation}\label{2.10}
   S(a(t),
b(t),\bar{p},\bar{q},\bar{p'},\bar{q'}) = S(a(t), b(t)
,\bar{p'},\bar{q'},\bar{p},\bar{q})
\end{equation}
where $C_{1} > 0$ is a given constant.
\\   6)
As consequences of the conservation law $p + q = p' + q'$, that
gives :
\begin{equation}\label{eq:2.11}
p^{o} + q^{o} = (p')^{o} + (q')^{o}\, ; \quad \bar{p} + \bar{q} =
\bar{p'} + \bar{q'}
\end{equation}

\begin{itemize}
  \item [i)]We have, using (2.4) and the first equation
  (2.11) :
\begin{equation}\label{eq:2.12}
 \sqrt{1 + \mid \bar{p} \mid^{2}} +  \sqrt{1 + \mid \bar{q} \mid^{2}} =
  \sqrt{1 + \mid \bar{p}' \mid^{2}} +  \sqrt{1 + \mid \bar{q}' \mid^{2}}
\end{equation}
which expresses the conservation of the quantity :
\begin{equation}\label{eq:2.13}
e = \sqrt{1 + \mid \bar{p} \mid^{2}} +  \sqrt{1 + \mid \bar{q}
\mid^{2}}
\end{equation}
called the energy of the unit rest mass particles.
\item[ii)]
We set, to express the second equation (\ref{eq:2.11}) and
following Glassey, R. T., in \cite{glassey2} :
\begin{equation}\label{eq:2.14}\begin{cases}
\bar{p'} = \bar{p} + C(\bar{p}, \bar{q}, \omega)\omega\\
\bar{q'} = \bar{q} - C(\bar{p}, \bar{q}, \omega)\omega
\end{cases}
\end{equation}
in which $C(\bar{p}, \bar{q}, \omega)$ is a real-valued function .
The relations (\ref{eq:2.12}) and (\ref{eq:2.14}) lead to a
quadratic equation in $C$ that solves to give the only non-trivial
solution :
\begin{equation}\label{eq:2.15}
C(\bar{p}, \bar{q}, \omega)\, = \, \frac{2\, p^{o} q^{o}e\, \,
\omega . (\hat{\bar{q}} - \hat{\bar{p}})}
 {e^{2} - [\omega . (\bar{p} + \bar{q})]^{2}}
\end{equation}
with $\hat{\bar{p}} = \frac{\bar{p}}{p^{o}}$, $e$ defined by
(\ref{eq:2.13}) and the dot product by (\ref{eq:2.3}). Now, using
the usual properties of determinants, the Jacobian of the change
of variables $(\bar{p}, \bar{q}) \longmapsto (\bar{p'}, \bar{q'})$
defined by (\ref{eq:2.14}) is computed to be :
\begin{equation}\label{eq:2.16}
\frac{\partial(\bar{p'}, \bar{q'})}{\partial(\bar{p}, \bar{q})} =
- \frac{p'^{o} q'^{o}}{p^{o} q^{o}}
\end{equation}
Notice that formulae (\ref{eq:2.15}) and (\ref{eq:2.16}) are
generalizations to the considered Bianchi type I space-time, of
analogous formulae established in Glassey, R. T.,
\cite{glassey2}, in the case of the Minkowski space-time, whose
metric corresponds to the particular case of (\ref{eq:2.1}), when
$a(t) = b(t) = 1$.
\end{itemize}
It then appears clearly, using (\ref{eq:2.4}) and (\ref{eq:2.14}),
that the functions in the integrals (\ref{eq:2.7}) and (\ref{2.8})
depend only on $\bar{p}, \bar{q}, \omega$ and that these integrals
give functions $Q^{+}(f, g) ,\,  Q^{-}(f, g )$, of the single
variable $\bar{p}$. In practice, we will consider functions $f$ on
$\mathbb{R}^{4}$ that induce for $t \in \mathbb{R}$, functions
$f(t)$ on $\mathbb{R}^{3}$, defined by : $f(t)(\bar{p}) = f(t,
\bar{p})$.\\ Now solving the Boltzmann equation (\ref{eq:2.5}),
which is a first order p.d.e. , is equivalent to solving the
associated characteristic system, that writes, taking $t$ as
parameter :
\begin{equation}\label{eq:2.17}
\frac{dp^{1}}{dt} =  -2 \,
\frac{\dot{a}}{a};\quad\frac{dp^{j}}{dt} = -2 \,
\frac{\dot{b}}{b},\quad  j=2, 3 \, ; \quad\frac{df}{dt} =
\frac{1}{p^{o}} \, Q(f, f)
\end{equation}
We study the initial value problem for the system (\ref{eq:2.17}),
with the initial data :
\begin{equation}\label{eq:2.18}
  p^{i}(0) = y^{i} ; \qquad f(0) = f_{o}
\end{equation}
In (\ref{eq:2.17}), the equations in $p^{i}$ integrate at once to
give, setting  $y = (y^{i}) \in \mathbb{R}^{3}$ :
\begin{equation}\label{eq:2.19}
\bar{p}(t, y) = A(t) y, \, \text{with} : \quad A(t) = \text{Diag}
\left( \frac{a^{2}(0)}{a^{2}(t)}, \quad \frac{b^{2}(0)}{b^{2}(t)},
\quad \frac{b^{2}(0)}{b^{2}(t)} \right)
\end{equation}
whereas the initial value problem for $f$ is equivalent to the
integral equation :
\begin{equation}\label{eq:2.20}
  f(t, y) = f_{o}(y) + \int_{0}^{t}\frac{1}{p^{o}} Q\big(f, f\big)(s, y)\,ds
\end{equation}
Finally, solving the initial value problem for the relativistic
Boltzmann equation (\ref{eq:2.5}), with the initial data $f_{o}$,
is equivalent to solving the integral equation (\ref{eq:2.20}) in
$f$. The strategy we adopt to solve (\ref{eq:2.20}) will consist
of :
\begin{itemize}
  \item [a)]Constructing approximating operators $Q_{n}$, with
  suitable properties, that will converge in the $L^{1}(\mathbb{R}^{3})$-norm
to the operator $\frac{1}{p^{o}}\,Q$.
  \item [b)]Solving by usual methods, the following approximating
  integral equation:
\begin{equation}\label{eq:2.21}
  f(t, y) = f_{o}(y) + \int_{0}^{t}Q_{n}\big(f, f\big)(s, y)\,ds
\end{equation}
obtained by replacing in (\ref{eq:2.20}), $\frac{1}{p^{o}}\,Q$ by
$Q_{n}$, and to obtain a global solution $f_{n}$ that will
converge to a global solution $f$ of (\ref{eq:2.20}), in a
suitable function space. Notice that, in $\mathbb{R}^{3}$, by
(\ref{eq:2.19}), the volume elements $d\bar{p}$, $dy$, and the
corresponding norms
 $\parallel . \parallel_ {\bar{p}} $, and $\parallel . \parallel_ {y} $ we
adopt to be $\parallel . \parallel$ , are linked by :
\begin{equation}\label{eq:2.22}
d\bar{p} = \frac{a^{2}(0) b^{4}(0)}{a^{2}(t) b^{4}(t)}dy \, ;\quad
\parallel . \parallel_ {\bar{p}}\, =\, \frac{a^{2}(0) b^{4}(0)}{a^{2}(t) b^{4}(t)}
 \parallel . \parallel
\end{equation}
\end{itemize}
\section{Preliminary results and approximating \\operators.}
In this section, we give properties of the collision operator $Q$,
we construct and we give the properties of a sequence of
approximating operators $Q_{n}$ to $\frac{1}{p_{o}}\,Q$.
\begin{proposition} \label{P:3.1}:\\
If $f, g \in L^{1}(\mathbb{R}^{3})$, then \mbox{$Q^{+}(f, g),
Q^{-}(f, g) \in  L^{1}(\mathbb{R}^{3})$}, and;
\begin{eqnarray}
\parallel \frac{1}{p^{o}} Q^{+}(f, g) \parallel & \leq &C \parallel f \parallel \,
\parallel g \parallel \\
\parallel \frac{1}{p^{o}} Q^{-}(f, g) \parallel & \leq &C \parallel f \parallel \,
\parallel g \parallel \\
\parallel \frac{1}{p^{o}} Q^{+}(f, f) - \frac{1}{p^{o}} Q^{+}(g, g)
\parallel & \leq & C\left( \parallel f \parallel + \parallel g \parallel \right)
\parallel f - g \parallel \\
\parallel \frac{1}{p^{o}} Q^{-}(f, f) - \frac{1}{p^{o}} Q^{-}(g, g)
\parallel & \leq & C\left( \parallel f \parallel + \parallel g \parallel \right)
\parallel f - g \parallel \\
\parallel \frac{1}{p^{o}} Q(f, f) - \frac{1}{p^{o}} Q(g, g)
\parallel & \leq & C\left( \parallel f \parallel + \parallel g \parallel \right)
\parallel f - g \parallel
\end{eqnarray}
where $ C = 8 \pi C_{1} a^{2}(0) b^{4}(0) \diagup C_{o}^{3} $
\end{proposition}
\textbf{Proof}
\begin{itemize}
\item [1)] The expression (\ref{eq:2.7}) for $Q^{+}(f, g)$
gives, using the upper bound (\ref{eq:2.9}) of S
\begin{equation*}
\parallel\frac{1}{p^{o}}\,Q^{+}(f, g) \parallel_ {\bar{p}} \leq C_{1} a(t)
b^{2}(t)
\int_{\mathbb{R}^{3}}\,\int_{\mathbb{R}^{3}}\,\frac{d\bar{p}d\bar{q}}{p^{o}q^{o}}
\int_{S^{2}}\mid f(\bar{p'}) \mid \,\mid g(\bar{q'}) \mid \,
d\omega \qquad  (a)
\end{equation*}
Now, the change of variables $ (\bar{p}, \bar{q}) \longmapsto
(\bar{p'}, \bar{q'})$, defined by (\ref{eq:2.14}) and its Jacobian
(\ref{eq:2.16}) give $ d\bar{p}\, d\bar{q} = -\frac{p^{o}
q^{o}}{p'^{o}\,q'^{o}}d\bar{p'}\,d\bar{q'}$; then we have from
(a), using (\ref{eq:2.4}) that gives $\frac{1}{(p')^{o}\,(q'^{o})}
\leq \,1 $:
\begin{equation*}
\parallel \frac{1}{p^{o}}\,Q^{+}(f, g) \parallel_ {\bar{p}} \leq C_{1} a(t)
b^{2}(t) \int_{\mathbb{R}^{3}}\,\parallel f(\bar{p'})\parallel
\,d\bar{p'}\int_{\mathbb{R}^{3}}\,\parallel g(\bar{q'})\parallel
\,d\bar{q'}\int_{S^{2}}d\omega = M
\end{equation*}
where
\begin{equation*}
M = 4 \pi C_{1} a(t)b^{2}(t)\parallel f
\parallel_{\bar{p}}\,\parallel g
\parallel_{\bar{p}}
\end{equation*}
and (3.1) then follows from the second relation (\ref{eq:2.22})
and the lower bound (\ref{eq:2.2})for $a$ and $b$ $\Box$
\item [2)] The expression (2.8) for $Q^{-}(f, g)$ gives,
using (\ref{eq:2.9}) and (\ref{eq:2.4}) that gives
$\frac{1}{p^{o}\,q^{o}} \leq \,1 $:
\begin{equation*}
\parallel \frac{1}{p^{o}}\,Q^{-}(f, g) \parallel_ {\bar{p}} \leq C_{1} a(t)
b^{2}(t) \int_{\mathbb{R}^{3}}\,\parallel f(\bar{p})\parallel
\,d\bar{p}\int_{\mathbb{R}^{3}}\,\parallel g(\bar{q})\parallel
\,d\bar{q}\int_{S^{2}}d\omega = M
\end{equation*}
and (3.2) then follows as above. $\Box$
\item[3)] Inequalities (3.3), (3.4), (3.5)
are direct consequences of (3.1), (3.2),\\$Q = Q^{+} - Q^{-}$, and
the fact that, the expressions (\ref{eq:2.7}) and (2.8) show that
$Q^{+}$ and $Q^{-}$ are bilinear operators, and so, we can write,
$P$ standing for $\frac{1}{p^{o}}\, Q^{+}$ or $\frac{1}{p^{o}}\,
Q^{-}$ :
\begin{equation*}
P(f, f) - P(g, g) =  P(f, f - g) + P(f - g, g)
\end{equation*}
This completes the proof of Proposition 3.1  $\Box$
\end{itemize}
We now state and prove the following result on which relies the
construction of the approximating operators; $C$ denotes the
constant defined in Proposition 3.1
\begin{proposition} \label{P:3.1}
Let $r$ be an arbitrary strictly positive real number. Then, there
exists an integer $n_{o}(r) > 0$ such that, for every integer $n
\geq n_{o}(r)$ and for every $v \in X_{r}$, the equation :
\begin{equation}\label{eq:3.6}
\sqrt{n} u \,- \,\frac{1}{p^{o}} Q(u, u) \,= \, v
\end{equation}
has a unique solution $u_{n} \in X_{r}$
\end{proposition}
\textbf{Proof}\\
Let $ r > 0$ be given and let $ v \in X_{r}$. We can write (3.6),
using $Q = Q^{+} - Q^{-}$ :
\begin{equation*}
\sqrt{n} u \,- \,\frac{1}{p^{o}} Q^{+}(u, u) + \,\frac{1}{p^{o}}
Q^{-}(u, u) \,= \, v
\end{equation*}
Notice that one deduces from the expression (2.8) of $Q^{-}$,
since $\bar{p}$ is not a variable for that integral, that
 $Q^{-}(f, g) = f Q^{-}(1, g)$. Hence the above equation can be
written, using $Q^{-}(u, u) = u Q^{-}(1, u)$, for $u \geq 0$ a.e.
\begin{equation*}
  u \, = \, F_{n}(u) \tag{a}
\end{equation*}
where
\begin{equation*}
F_{n}(h) = \frac{v + \frac{1}{p^{o}} Q^{+}(h, h)}{\sqrt{n} +
\frac{1}{p^{o}} Q^{-}(1, h)} \tag{b}
\end{equation*}
(a) shows that a solution $u$ of (3.6) can be considered as a
fixed point of the map $h \longmapsto F_{n}(h)$ defined by (b).
Let us shows that  $F_{n}$ is a contracting map from $X_{r}$ to
$X_{r}$. We will then solve the problem by applying the fixed
point theorem .
\begin{itemize}
\item [i)]We deduce from (3.1) that $F_{n} : X_{r} \longmapsto
L^{1}(\mathbb{R}^{3})$ , and that since $v , h \in X_{r}$ :
\begin{equation*}
\parallel F_{n}(h)\parallel \, \leq \, \frac{\parallel v
\parallel}{\sqrt{n}}\, + \, \frac{C\parallel h \parallel \,\parallel h
\parallel}{\sqrt{n}}\, \leq \,\frac{r(1+Cr)}{\sqrt{n}} \tag{c}
\end{equation*}
\item[ii)] Let $g,\, h \in X_{r}$ . (b) gives :
\begin{equation*}
F_{n}(h) - F_{n}(g) = \frac{v + \frac{1}{p^{o}} Q^{+}(h,
h)}{\sqrt{n} + \frac{1}{p^{o}} Q^{-}(1, h)} - \frac{v +
\frac{1}{p^{o}} Q^{+}(g, g)}{\sqrt{n} + \frac{1}{p^{o} }Q^{-}(1,
g)} \tag{d}
\end{equation*}
Since $g \geq 0$, $\,h \geq 0$ a.e., we have , taken $n \geq 1$:
\begin{equation*}
  [\sqrt{n} + \frac{1}{p^{o}} Q^{-}(1, h)][\sqrt{n} + \frac{1}{p^{o}} Q^{-}(1, g)]
\, \geq \, n \geq \sqrt{n}.
\end{equation*}
We then deduce from (d), using once more $Q^{-}(f, g) = fQ^{-} (1,
g )$, that :
\begin{align*}
\sqrt{n} \parallel F_{n}(h) - F_{n}(g) \parallel &\leq
\parallel \frac{1}{p^{o}} Q^{+}(h, h) - \frac{1}{p^{o}} Q^{+}(g, g) \parallel
+ \parallel \frac{1}{p^{o}} Q^{-}(v, g) - \frac{1}{p^{o}} Q^{-}(v,
h) \parallel\\
& \qquad  + \parallel \frac{1}{p^{o}} Q^{-}[\frac{1}{p^{o}}
Q^{+}(h, h), g] - \frac{1}{p^{o}} Q^{-}[\frac{1}{p^{o}} Q^{+}(g,
g), h] \parallel \qquad  (e)
\end{align*}
\end{itemize}
We can write, using the bilinearity of $Q^{-}$ :
\begin{equation*}
\frac{1}{p^{o}}\, Q^{-}(v, g) - \frac{1}{p^{o}}\, Q^{-}(v, h) =
\frac{1}{p^{o}}\, Q^{-}(v, g - h)
\end{equation*}
So, in (e) we apply (3.3) to the first term, (3.2) to the second
term to obtain, using $\parallel g \parallel \leq r$, $\parallel h
\parallel \leq r$, $\parallel v \parallel \leq r$ :
\begin{equation*}
\parallel \frac{1}{p^{o}}\, Q^{+}(h, h) - \frac{1}{p^{o}}\, Q^{+}(g, g) \parallel
 + \parallel \frac{1}{p^{o}}\, Q^{-}(v, g) + \frac{1}{p^{o}}\,Q^{-}(v, h)
\parallel \leq 3Cr\parallel h - g \tag{f}
\end{equation*}
Concerning the last term in (e), we can write, using the
bilinearity of $Q^{-}$ :
\begin{align*}
\frac{1}{p^{o}}\, Q^{-}\left[\frac{1}{p^{o}}\,Q^{+}(h, h), g
\right] - \frac{1}{p^{o}}\,Q^{-}\left[\frac{1}{p^{o}}\, Q^{+}(g,
g), h \right] &= \frac{1}{p^{o}}\, Q^{-}\left[\frac{1}{p^{o}}\,
Q^{+}(h, h),
g -h \right]\\
& \qquad  +  \frac{1}{p^{o}}\, Q^{-}\left[\frac{1}{p^{o}}\,
Q^{+}(h, h) - \frac{1}{p^{o}}\,Q^{+}(g, g), h \right]
\end{align*}
We then apply to the right hand side of the above equality :
\begin{itemize}
  \item [1)] For the first term, property (3.2) of $Q^{-}$,
followed by property (3.1) of $Q^{+}$
  \item[2)] For the second term, property (3.2) of $Q^{-}$,
followed by property (3.3) of $Q^{+}$ to obtain, using $\parallel
g \parallel \leq r$, $\parallel h \parallel \leq r$ :
\begin{equation*}
\parallel \frac{1}{p^{o}} Q^{-}\left[\frac{1}{p^{o}} Q^{+}(h, h), g \right]
- \frac{1}{p^{o}} Q^{-}\left[\frac{1}{p^{o}} Q^{+}(g, g), h
\right]
\parallel \, \leq \, 3C^{2}r^{2}\parallel h - g \parallel  \tag{h}
\end{equation*}
thus (e) gives, using (f) and (h) :
\begin{equation*}
 \parallel F_{n}(h) - F_{n}(g) \parallel \leq \frac{3Cr(1 + Cr)}{\sqrt{n}}
\parallel h - g \parallel \tag{i}
\end{equation*}
\end{itemize}
It is then easy , using (c) and (i), to choose an interger
$n_{o}(r)$ such that, for $n \geq n_{o}(r)$ , we have:
\begin{equation}\label{eq:3.7}
\frac{1 + Cr}{\sqrt{n}} \leq 1 \quad and \quad \frac{3Cr(1 +
Cr)}{\sqrt{n}} \leq \frac{1}{2}
\end{equation}
This shows , using once more (c) and (i) that, for $n \geq
n_{0}(r)$, $F_{n}$ is a contracting map of the complete metric
space $X_{r}$ into itself. By the fixed point theorem, $F_{n}$ as
a unique fixed point $u_{n} \in X_{r}$ that is the unique solution
of (3.6) $\Box$\\
 In all the following, $n_{o}(r)$ stands for the
integer defined above.\\
 The above result leads us to the following definition :
\begin{definition}\label{d:3.1}
 Let $n \in
\mathbb{N}$, $n \geq n_{o}(r)$ be given.
\begin{itemize}
  \item [1)] Define the operator
\begin{equation*}
R(n, Q) : X_{r} \longrightarrow X_{r},\quad u \longmapsto R(n, Q)u
\end{equation*}
as follows : for $u \in X_{r}$, $R(n, Q)u$ is the unique element
of $X_{r}$ such that :
\begin{equation}\label{eq:3.8}
 \sqrt{n}R(n, Q)u - \frac{1}{p^{o}} Q\left[R(n, Q)u, R(n, Q)u\right] = u
\end{equation}
  \item[2)] Define the operator $Q_{n}$ on $X_{r}$ by :
\begin{equation}\label{eq:3.9}
  Q_{n}(u, u) = n\sqrt{n}R(n, Q)u - nu
\end{equation}
 \end{itemize}
\end{definition}
We give the properties of the operators $R(n, Q)$ and $Q_{n}$.
\begin{proposition}\label{p:3.3}:
Let $r \in \mathbb{R}_{+}^{*}$, $n, m \in \mathbb{N}$ $m \geq
n_{o}(r)$, $n \geq n_{o}(r)$, be given.Then we have, for every $u,
v \in X_{r}$
\begin{eqnarray}
\parallel \sqrt{n} R(n, Q)u \parallel &  = & \parallel u \parallel \\
\parallel \sqrt{n} R(n, Q)u \,- \, u \parallel & \leq & \frac{K}{n} \\
\parallel R(n, Q)u \, - \,R(n, Q)v \parallel &  \leq &
\frac{2}{\sqrt{n}} \parallel u - v \parallel \\
Q_{n}(u, u) & = & \frac{1}{p^{o}} Q\left[ \sqrt{n} R(n, Q)u,
\sqrt{n} R(n, Q)u
\right] \\
\parallel Q_{n}(u, u) - Q_{n}(v, v) \parallel & \leq & K\parallel u - v \parallel
\\
\parallel Q_{n}(u, u) - \frac{1}{p^{o}} Q(v, v) \parallel & \leq &
\frac{K}{n} +  K\parallel u - v \parallel \\
\parallel Q_{n}(u, u) - Q_{m}(v, v) \parallel & \leq &
K\parallel u - v \parallel + \frac{K}{n} + \frac{K}{m}
\end{eqnarray}
Here $K = K(r)$ is a continuous function of $r$.
\end{proposition}
\textbf{Proof}. \\
1) (3.10) will be a consequence of the relation:
\begin{equation}\label{eq:3.17}
\int_{\mathbb{R}^{3}}\,\frac{1}{p^{o}} Q(f, g)\,d\bar{p} = 0,
\qquad \forall f, g \in L^{1} (\mathbb{R}^{3})
\end{equation}
we now establish. This is where we need assumption (2.10) on the
kernel $S$. We have, using  definition (2.7) of $Q^{+}$, the
change of variables $(\bar{p}, \bar{q}) \longmapsto
(\bar{p'}\bar{q'})$ defined by (2.14) and its Jacobian (2.16), and
since $ p^{o} = \sqrt{1 + \mid \bar{p} \mid^{2}}$ :
\begin{equation*}
I_{o} = \int_{\mathbb{R}^{3}}\,\frac{1}{p^{o}} Q^{+}(f,
g)\,d\bar{p} = a(t) b^{2}(t)\int_{\mathbb{R}^{3}}\,d\bar{p'}
\int_{\mathbb{R}^{3}}\,d\bar{q'}\int_{S^{2}}\,\frac{f(\bar{p'})
g(\bar{q'}) S(\bar{p'}, \bar{q'}, \bar{p}, \bar{q}) }{\sqrt{1 +
\mid \bar{p'} \mid^{2}}\,\sqrt{1 + \mid \bar{q'}
\mid^{2}}}\,d\omega
\end{equation*}
On the other hand, we have, using definition (2.8) of $Q^{-}$ and
$ p^{o} = \sqrt{1 + \mid \bar{p} \mid^{2}}$ :
\begin{equation*}
J_{o} = \int_{\mathbb{R}^{3}}\,\frac{1}{p^{o}} Q^{-}(f,
g)\,d\bar{p} = a(t) b^{2}(t)\int_{\mathbb{R}^{3}}\,d\bar{p}
\int_{\mathbb{R}^{3}}\,d\bar{q}\int_{S^{2}}\,\frac{f(\bar{p})
g(\bar{q}) S(\bar{p}, \bar{q}, \bar{p'}, \bar{q'}) }{\sqrt{1 +
\mid \bar{p} \mid^{2}}\,\sqrt{1 + \mid \bar{q} \mid^{2}}}\,d\omega
\end{equation*}
It then appears that $I_{o} = J_{o}$ and (\ref{eq:3.17}) follows
from $Q = Q^{+} - Q^{-}$. Then, to obtain (3.10), we integrate
(3.8) over $\mathbb{R}^{3}$ with respect to $\bar{p}$, use (3.17),
and conclude by (\ref{eq:2.22}) that gives the equivalence between
$\parallel . \parallel_ {\bar{p}}$ and
$\parallel . \parallel$ $\Box$ \\
2) The definition (\ref{eq:3.8}) of $R(n, Q)$ gives, using
property (3.5) of $Q$ with \\$f =  R(n, Q)u$ and $g = 0$ :
\begin{equation*}
  \parallel \sqrt{n} R(n, Q)u - u \parallel = \parallel\frac{1}{p^{o}}
Q\left[R(n, Q)u, R(n, Q)u\right]\parallel \leq C
\parallel R(n, Q)u \parallel^{2}
\end{equation*}
and (3.11) follows from (3.10) that gives $\parallel R(n,
Q)u\parallel = \frac{\parallel u \parallel}{\sqrt{n}} \leq
\frac{r}{\sqrt{n}}$.\\
 3) Subtracting the relations (3.8) written for $u$ and $v$ gives:
\begin{equation*}
R(n, Q)u - R(n, Q)v = \frac{u - v}{\sqrt{n}} +
\frac{1}{\sqrt{n}p^{o}} Q\left[R(n, Q)u, R(n, Q)u\right] -
\frac{1}{\sqrt{n}p^{o}} Q\left[R(n, Q)v, R(n, Q)v\right]
\end{equation*}
which gives, using property (3.5) of $Q$ and once more (3.10):
\begin{equation*}
\parallel R(n, Q)u - R(n, Q)v \parallel \leq \frac{\parallel u - v \parallel}
{\sqrt{n}} + \frac{2Cr}{\sqrt{n}} \parallel R(n, Q)u - R(n, Q)v
\parallel
\end{equation*}
(3.12) then follows from $n \geq n_{o}(r)$ and (3.7) that implies
$\frac{2Cr}{\sqrt{n}} \leq \frac{1}{2}$.\\
4) The expression (3.13) of $Q_{n}$ is obtained by multiplying
(3.8) by $n$, using its definition (3.9) and the bilinearity of
$Q$.\\
5) Property (3.14) is obtained by using (3.13), property (3.5) of
$Q$, followed by properties (3.10) and (3.12) of $R(n, Q)$. \\
6) Notice that by (3.10), $\sqrt{n} R(n, Q)u \in X_{r}$ since $u
\in X_{r}$. Then, the expression (3.13) of $Q_{n}$ and property
(3.5) of $Q$ give :
\begin{equation*}
\parallel Q_{n}(u, u) - \frac{1}{p^{o}} Q(v, v) \parallel \leq 2Cr
\parallel \sqrt{n} R(n, Q)u - v \parallel \leq
2Cr\left(\parallel \sqrt{n} R(n, Q)u - u\parallel +
\parallel u - v \parallel\right)
\end{equation*}
hence, (3.15) follows from (3.11). \\
7) The expression (3.13) of $Q_{n}$ gives:
\begin{equation*}
Q_{n}(u, u) - Q_{m}(v, v) =  \frac{1}{p^{o}} Q\left[ \sqrt{n} R(n,
Q)u, \sqrt{n} R(n, Q)u \right] - \frac{1}{p^{o}} Q\left[ \sqrt{m}
R(m, Q)v, \sqrt{m} R(m, Q)v
\right] \\
\end{equation*}
then, (3.16) follows from (3.5), (3.10) and (3.11), adding and
subtracting $u$ and $v$. $\Box$

\begin{remark}.
(3.15) shows by taking $u = v$, that $Q_{n}$ converges pointwise
in the $L^{1}(\mathbb{R}^{3})$-norm to the operator
$\frac{1}{p^{o}} Q$. From there, the qualification of \\
''approximating operators'' we give to the sequence $Q_{n}$.
\end{remark}
We now have all the tools we need to prove the global existence
theorem.
\section{The global existence theorem}
With the approximating operator $Q_{n}$ defined by (3.8), we will
first state and prove a global existence theorem for the
''approximating'' integral equation:
\begin{equation}\label{eq:4.1}
  f(t_{o} + t, y) = g_{o}(y) + \int_{t_{o}}^{t_{o} +
t}Q_{n}\left(f, f\right)(s, y)\,dy , \qquad t \geq 0,
\end{equation}
in which $t_{o}$ is an arbitrary positive real number, and $g_{o}$
is a given function that stands for the initial data at time
$t_{o}$; for practical reasons,we study (\ref{eq:4.1}) rather than
(\ref{eq:2.21}) that is the particular case when $t_{o} = 0$ and
$g_{o} = f_{o}$. Next, we prove, by a convenient choice of $t_{o}$
that the global solution $f_{n}$ of (\ref{eq:4.1}) converges in a
sense to specify, to a global solution $f$ of (\ref{eq:2.20}).
\begin{proposition}\label{p:4.1}
Let $r$ be an arbitrary strictly positive real number.\\
 Let $g_{o} \in X_{r}$ and let $n$ be an integer such that $n \geq
n_{o}(r)$. \\
Then, for every $t_{o} \in [0, +\infty[$, the
integral equation (\ref{eq:4.1}) has a unique solution $f_{n} \in
C[t_{o}, +\infty; X_{r}]$. Moreover, $f_{n}$ satisfies the
inequality :
\begin{equation}\label{eq:4.2}
\mid\parallel f_{n} \mid\parallel \leq \parallel g_{o} \parallel
\end{equation}
\end{proposition}
\textbf{Proof}: \\
We give the proof in 2 steps. \\
\textbf{Step 1}:Local existence and estimation. \\
Consider equation (\ref{eq:4.1}) for $t \in [t_{o}, t_{o} +
\delta]$ where $\delta  >0$ is given. One verifies easily, using
(3.8) that gives : $ Q_{n}(f, f) + n f = n\sqrt{n}R(n, Q)f $  and
\\$ \frac{d}{dt} [e^{nt}f(t_{o} + t, y)] = e^{nt}\left[
\frac{df}{dt} + nf \right](t_{o} + t, y) $, that (\ref{eq:4.1}) is
equivalent to :
\begin{equation}\label{eq:4.3}
 f(t_{o} + t, y) = e^{-nt}g_{o}(y) + \int_{0}^{
t} n\sqrt{n} e^{-n(t - s)}R(n, Q)f(t_{o} + s, y)\,dy
\end{equation}
We solve (\ref{eq:4.3}) by the fixed point theorem. Let us define
the operator $A$ over the complete metric space $C[t_{o}, t_{o} +
\delta; X_{r}]$ by :
\begin{equation}\label{eq:4.4}
Af(t_{o} + t, y) = e^{-nt}g_{o}(y) + \int_{0}^{ t}n\sqrt{n}
e^{-n(t - s)}R(n, Q)f(t_{o} + s, y)\,dy
\end{equation}
\begin{itemize}
  \item [i)] Let $f \in C[t_{o}, t_{o} + \delta; X_{r}]$;
(\ref{eq:4.4}) gives, since $\parallel g_{o}\parallel \leq r$ and
using (3.9) that gives, $\parallel \sqrt{n} R(n, Q)f(t_{o} +
s)\parallel = \parallel f(t_{o} + s)\parallel\, \leq r $, and for
every $t \in [0, \delta]$ :\begin{equation*}
\parallel Af(t_{o} + t) \parallel\, \leq \,e^{-nt}r + nre^{-nt}\int_{0}^{t}
e^{ns}\,ds = e^{-nt}r + re^{-nt}(e^{nt} - 1) = r
\end{equation*}
Hence $\mid\parallel Af \mid\parallel \leq r$, and this shows that
$A$ maps $C[t_{o}, t_{o} + \delta; X_{r}]$ into itself.
\item[ii)] Let $f, g \in C[t_{o}, t_{o} + \delta; X_{r}]$; $t \in [0,
\delta]$; (4.4)gives:
\begin{equation*}
 \left(Af - Ag\right)(t_{o} + t, y) = \int_{0}^{ t} n\sqrt{n} e^{-n(t -
s)}\left[R(n, Q)f - R(n, Q)g\right](t_{o} + s, y)\,dy
\end{equation*}
which gives, using property (3.12) of $R(n, Q)$ and $e^{-n(t - s)}
\leq 1$ :
\begin{equation*}
\mid\parallel Af - Ag \mid\parallel \leq
2n\delta\mid\parallel f - g \mid\parallel.
\end{equation*}
 \end{itemize}
It then appears that $A$ is a contracting map in a metric space
$C[t_{o}, t_{o} + \delta; X_{r}]$ where \, $2\,n\,\delta \leq
\frac{1}{2}$, i.e. $\delta \in ]0, \frac{1}{4n}]$. Taking $\delta
= \frac{1}{4n} $, we conclude that $A$ has a unique fixed point
  $f_{n}^{o} \in C[t_{o}, t_{o} + \frac{1}{4n}; X_{r}]$, that is
the unique solution of (4.3) and hence, of (4.1). \\
Now we have, by taking $t = 0$ in (4.3), $f(t_{o}, y) = g_{o}(y)$.
The solution $f_{n}^{o}$ then satisfies:
\begin{equation*}
 f_{n}^{o}(t_{o} + t, y) = e^{-nt}f_{n}^{o}(t_{o}, y) + \int_{0}^{
t} n\sqrt{n} e^{-n(t - s)}R(n, Q)f_{n}^{o}(t_{o} + s, y)\,dy
\end{equation*}
which gives, multiplying by $e^{nt}$, using once more (3.10), and
for $t \in ]0, \frac{1}{4n}]$:
\begin{equation*}
 \parallel e^{nt}f_{n}^{o}(t_{o} + t) \parallel\, \leq \, \parallel f_{n}^{o} \parallel
 + n\int_{0}^{t}\,\parallel e^{ns}f_{n}^{o}(t_{o} + s) \parallel \,ds
\end{equation*}
this gives, by Gronwall Lemma: $e^{nt}\parallel f_{n}^{o}(t_{o} +
t )\parallel \leq e^{nt}\parallel f_{n}^{o}(t_{o})\parallel$ and
hence
\begin{equation}\label{eq:4.5}
 \mid\parallel f_{n}^{o} \mid\parallel\, \leq \, \parallel f_{n}^{o}(t_{o}) \parallel
\end{equation}
\textbf{Step 2:} Global existence; estimation and uniqueness. \\
Let $k \in \mathbb{N}$. By replacing in the integral equation
(4.1) $t_{o}$ by : \\ \mbox{$t_{o} + \frac{1}{4n}, t_{o} +
\frac{2}{4n}, \cdots, t_{o} + \frac{k}{4n}, \cdots$}, step 1 tells
us that on each interval \\ $I_{k} = \left[t_{o} + \frac{k}{4n},
t_{o} + \frac{k}{4n} + \frac{1}{4n}\right]$ whose length is
$\frac{1}{4n}$, the initial value problem for the corresponding
integral equation has a unique solution $f_{n}^{k} \in C[I_{k},
X_{r}]$, provided that the initial data we denote $f_{n}^{k}\left[
t_{o} + \frac{k}{4n} \right]$ is a \textbf{given} element of
$X_{r}$; $f_{n}^{k}$ then satisfies:
\begin{equation*}
\begin{cases}
f_{n}^{k}(t_{o} + \frac{k}{4n} + t, y) = f_{n}^{k}(t_{o} +
\frac{k}{4n}, y) + \int_{t_{o} + \frac{k}{4n}}^{t_{o} +
\frac{k}{4n} + t}\, Q_{n}\left( f_{n}^{k},
f_{n}^{k}\right)(s,y)\,ds \\
f_{n}^{k}\left(t_{o} + \frac{k}{4n}\right) \in X_{r};\quad k \in
\mathbb{N}; \quad t \in \left[ 0, \frac{1}{4n}\right]
\end{cases}\end{equation*}
and (4.5)implies:
\begin{equation*}
  \mid\parallel f_{n}^{k} \mid\parallel \,\leq \,\parallel f_{n}^{k}(t_{o} +
 \frac{k}{4n}) \parallel.
\end{equation*}
Now, $[0, +\infty[ = \underset{k \in \mathbb{N}}{\bigcup} \,\left[
t_{o} + \frac{k}{4n}, t_{o} + \frac{k}{4n} + t \right]$; Define :
\begin{equation*}
f_{n}^{k}(t_{o} + \frac{k}{4n}, y) = f_{n}^{k - 1}(t_{o} +
\frac{k}{4n}, y)\quad \text{if} \quad k \geq 1 \, \text{and} \quad
f_{n}^{o}(t_{o}) = g_{o}.
\end{equation*}
Then, the solution $f_{n}^{k - 1}$ and $f_{n}^{k}$ that are
defined on $\left[  t_{o} + \frac{k - 1}{4n},  t_{o} +
\frac{k}{4n} \right]$ and $\left[  t_{o} + \frac{k}{4n},  t_{o} +
\frac{k + 1}{4n} \right]$ overlap at $t =  t_{o} + \frac{k}{4n}$.\\
Define the function $f_{n}$ on $[0, +\infty[$ by
\begin{equation*}
 f_{n}(t) = f_{n}^{k}(t) \quad \text{if} \quad t \in \left[  t_{o} +
\frac{k}{4n},  t_{o} + \frac{k + 1}{4n} \right].
\end{equation*}
Then, a straightforward calculation shows, using the above
relations for\\ \mbox{$k, k - 1, k - 2, \cdots, 1,0,$} that
$f_{n}$ is a global solution of (4.1) on $[0, +\infty[$ satisfying
the estimation (4.2) and hence, $f_{n} \in C[t_{o}, +\infty;
X_{r}]$.\\
Now suppose that $f, g \in C[t_{o}, +\infty; X_{r}]$ are 2
solutions of (4.1), then:
\begin{equation*}
\left( f - g \right)(t_{o} + t, y) = \int_{o}^{t}\left[ Q_{n}(f,
f) - Q_{n}(g, g)\right](t_{o} + s, y)\,ds \quad t \geq 0.
\end{equation*}
This gives, using property (3.14) of $Q_{n}$:
\begin{equation*}
 \parallel \left(f - g\right)(t_{o} + t)\parallel \,\leq \,K\int_{0}^{t}
\parallel \left(f - g\right)(t_{o} + s)\parallel \,ds \quad t \geq 0
\end{equation*}
which implies, using Gronwall Lemma, $f = g$, and the uniqueness
follows. This ends the proof of proposition 4.1.$\Box$

 We now state and prove the global existence theorem.
\begin{theorem}\label{T:4.1}
Let $f_{o} \in L^{1}(\mathbb{R}^{3})$, $f_{o} \geq 0$, $a.e$; be
arbitrary given.  Then the Cauchy problem for the relativistic
Boltzmann equation on the Bianchi type I space-time, with initial
data $f_{o}$, has a unique global solution $f \in C[0, +\infty;
L^{1}(\mathbb{R}^{3})]$, $f(t) \geq 0, a.e, \forall t \in [0,
+\infty[$; $f$ satisfies the estimation:
\begin{equation}\label{eq:4.6}
  \underset{t \in [0, +\infty[}{Sup}\parallel f(t)\parallel \,\leq
\, \parallel f_{o} \parallel
\end{equation}
\end{theorem}
\textbf{Proof}: \\
Fix a real number $r>\parallel f_{o} \parallel$. We give the proof in 2 steps. \\
 \textbf{\underline{Step 1}}: Local existence and estimation.

Let $t_{o} \geq 0$ and $g_{o} \in X_{r}$ be given. The proposition
4.1 gives for every integer $n \geq n_{o}(r)$, the existence of a
solution $f_{n} \in C[t_{o}, +\infty; X_{r}]$ of (4.1) that
satisfies the estimation (4.2). It is important to notice that
this solution depends on $n$. \\
Let $T > 0$ be given;  we have $f_{n} \in C[t_{o}, t_{o} + T;
X_{r}], \,\forall n \in \mathbb{N}, \, n \geq n_{o}(r)$. Let us
prove that the sequence $f_{n}$ converges in $ C[t_{o}, t_{o} + T;
X_{r}]$ to a solution of the integral equation :
\begin{equation}\label{eq:4.7}
f(t_{o} + t, y) = g_{o}(y) + \int_{t_{o}}^{t_{o} + t}
\frac{1}{p^{o}}Q\left(f, f\right)(s, y)\,dy , \qquad t \in[0, T]:
\end{equation}
Consider 2 integers $m, n \geq n_{o}(r)$ ; we deduce from (4.1),
that, $\forall t \in [0, T]$;
\begin{equation*}
 f_{n}(t_{o} + t, y) - f_{m}(t_{o} + t, y)=  \int_{t_{o}}^{t_{o} +
t}\left[Q_{n}\left(f_{n}, f_{n}\right) - Q_{m}\left(f_{m}, f_{m}
\right)\right](s, y)\,dy
\end{equation*}
this gives, using the property (3.16) of the operators $Q_{n}$,
$Q_{m}$ and Gronwall lemma:
\begin{equation*}
 \mid\parallel f_{n} - f_{m} \mid\parallel \,\leq \,\left(\frac{1}{n}
+ \frac{1}{m} \right)KTe^{KT}
\end{equation*}
this proves that $f_{n}$ is a Cauchy sequence in the complete
metric space \\$ C[t_{o}, t_{o} + T, X_{r}]$. Then, there exists
$f \in C[t_{o}, t_{o} + T, X_{r}]$ such that:
\begin{equation}\label{eq:4.8}
 f_{n} \,\text{ converges to }\,f \,\text{ in }\, C[t_{o}, t_{o} + T,
X_{r}]
\end{equation}
The definition of $X_{r}$ implies that $f(t_{o} + t) \geq 0 $
a.e $\forall t \in [0, T]$;\\
 Let us show that $f$ satisfies (4.7). The convergence (4.8)
implies that, $f_{n}(t_{o} + t)$ converges to $f(t_{o} + t),
\,\forall t \in [0, T]$; then $f_{n}(t_{o})$ converges to
$f(t_{o})$; but $f_{n}(t_{o}) = g_{o}$, then $f(t_{o}) = g_{o}$;
next we have, using property (3.14) of $Q_{n}$ and $Q$ :
\begin{equation*}
\parallel \int_{t_{o}}^{t_{o} +
t}\left[Q_{n}\left(f_{n}, f_{n}\right) - \frac{1}{p_{o}} Q\left(f,
f \right)\right](s, y)\,dy \parallel \,\leq \,\frac{KT}{n} +
KT\mid\parallel f_{n} - f \mid\parallel
\end{equation*}
which shows that
\begin{equation*}
\int_{t_{o}}^{t_{o} + t}Q_{n}\left( f_{n}, f_{n} \right)(s)\,ds\,
\text{ converges to }\,\int_{t_{o}}^{t_{o} + t}Q\left(f,
f\right)(s)\,ds \,\text{ in }\,L^{1}(\mathbb{R}^{3}).
\end{equation*}
Hence, since $f_{n}$ satisfies (4.1), $f$ satisfies (4.7),\,
$\forall t \in [0, T]$ and a.e. in $y \in \mathbb{R}^{3}$.
Finally, (4.2) and (4.8) imply:
\begin{equation}\label{eq:4.9}
  \mid\parallel f \mid\parallel \,\leq \,\parallel f(t_{o})\parallel
\end{equation}\\

\textbf{\underline{Step 2}}: Global existence, estimation and
uniqueness

Since $t_{o} \in [0, +\infty[$ is arbitrary and since $\forall n
\geq n_{o}(r)$, the solution $f_{n}$ of (4.1) is global on
$[t_{o}, +\infty[$, Step 1 tells us that, given $T > 0$, by taking
in the integral equation (4.7), $t_{o} = 0, t_{o} = T, t_{o} = 2T,
\cdots, t_{o} = kT, \cdots$ $k \in \mathbb{N}$, then, on each
interval $J_{k} = [kT, (k + 1)T]$, $k \in \mathbb{N}$, whose
length is $T$, the initial value problem for the corresponding
integral equation has a solution $f^{k} \in C[J_{k}, X_{r}]$,
provided that the initial data we denote $f^{k}(kT)$, is a given
element of $X_{r}$. \\
We then proceed exactly as in Step 2 of the proof of Proposition
4.1  by writing for $T > 0$ given, that $[0, +\infty[ =
\underset{k \in \mathbb{N}}{\bigcup}[kT, (k + 1)T]$, to overlap
the local solutions $f^{k}\in C[kT, (k + 1)T; X_{r}]$, by setting
$f^{k}(kT) = f^{k - 1}(kT)$ if $k \geq 1$ and $f^{o}(0) = f_{o}$,
and we obtain a global solution $f \in C[0, +\infty; X_{r}]$ of
the integral equation (2.20) defined by $f(t) = f^{k}(t)$, if $t
\in [kT, (k + 1)T]$. Next, (4.9) that gives
\begin{equation*}
  \mid\parallel f^{k}\mid\parallel \, \leq \,\parallel f^{k}(kT)\parallel
\end{equation*}
shows that $f$ satisfies the estimation (4.6). \\
Now if $f, g \in C[0, +\infty; L^{1}(\mathbb{R}^{3})]$ are 2
solutions of (2.20) for some initial data $f_{o}$, (4.6) and
property (3.5) of $Q$ give:
\begin{equation*}
\parallel\left(f - g\right)(t)\parallel \, \leq \,2C\parallel f_{o}\parallel\int_{0}^{t}
\parallel\left(f - g\right)(s)\parallel\,ds, \quad \forall t \geq
0.
\end{equation*}
which gives, by Gronwall Lemma, $f = g$, and the uniqueness
follows. \\Now as we indicated, solving the initial value problem
for the relativistic Boltzmann equation is equivalent to solving
the integral equation (2.20). This ends the proof of theorem
4.1$\Box$
\begin{remark}\label{r:4.1}
\begin{itemize}
\item[1)] The expression $C(\bar{p}, \bar{q}, \omega) = \omega.
(\bar{p} - \bar{q})$ used by the author in \cite{mucha1} is not
correct in the full relativistic case, and is to be replaced by
(2.15) in which the dot product that corresponds to the Bianchi
type I case is given by (2.3) . \item[2)] In \cite{mucha1} the
author defined the operator $R(n, Q)$ whose domain is $X_{r}$,
using the equation: $n u - \frac{1}{p^{o}} Q(u, u) = v$, instead
of (3.6), and he associated to it, the approximating operator
$P_{n}(u, u) = n R(n, Q)nu - nu $, instead of $Q_{n}$ defined by
(3.8). But it is clear that, even if $u \in X_{r}$, $nu$ is no
longer in $X_{r}$ for $n$ sufficiently large. So the domain of
$P_{n}$ is unspecified. \item[3)] One could use (3.5) that means
that $Q$ is locally lipschitzian in $f$, to solve the last
equation (2.17) using classical results on the differential
equation. But this method does not give the positivity property of
$f$, imposed by the physical nature of the considered problem, and
that we obtain by our method. \item[4)] In theorem 4.1, there is
no restriction on the size of $f_{o}$ which can be taken arbitrary
large in $L^{1}-norm$
\end{itemize}
\end{remark}
\textbf{\underline{Acknowledgement}}. The authors thank
A.D.Rendall for helpful comments and suggestions. This work was
supported by the VolkwagenStiftung, Federal Republic of Germany.

\end{document}